\documentclass[twocolumn,showpacs,preprintnumbers,amsmath,amssymb]{revtex4}
\usepackage[T1]{fontenc}
\usepackage[latin1]{inputenc}
\usepackage{amsmath}
\usepackage{graphicx}

\makeatletter

\newcommand{\lyxdot}{.}

\usepackage{epsfig}

\usepackage{dcolumn}
\usepackage{bm}
\usepackage{wasysym}

\usepackage{marvosym}

\usepackage{color}

\makeatletter

\usepackage{ulem}

\makeatother

\makeatother

\begin{document}

\title{Magnetic structure at zigzag edges of graphene bilayer ribbons}

\author{Eduardo V. Castro$^{1}$, N.M.R. Peres$^{2}$, J.M.B. Lopes dos Santos$^{1}$}

\affiliation{$^{1}$ CFP and Departamento de F\'{\i}sica, Faculdade de Ciências
Universidade do Porto, P-4169-007 Porto, Portugal}

\affiliation{$^{2}$Center of Physics and Departamento de F\'{\i}sica, Universidade
do Minho, P-4710-057, Braga, Portugal}

\begin{abstract}
We study the edge magnetization of bilayer graphene ribbons with zigzag
edges. The presence of flat edge-state bands at the Fermi energy of
undoped bilayer, which gives rise to a strong peak in the density
of states, makes bilayer ribbons magnetic at the edges even for very
small on-site electronic repulsion. Working with the Hubbard model
in the Hartree Fock approximation we show that the magnetic structure
in bilayer ribbons with zigzag edges is ferromagnetic along the edge,
involving sites of the two layers, and antiferromagnetic between opposite
edges. It is also shown that this magnetic structure is a consequence
of the nature of the edge states present in bilayer ribbons with zigzag
edges. Analogously to the monolayer case, edge site magnetization
as large as $m\approx0.2\mu_{B}$ (per lattice site) even at small
on-site Hubbard repulsion $U\approx0.3\,\mbox{eV}$ is realized in
nanometer wide bilayer ribbons.
\end{abstract}

\pacs{73.20.-r, 73.20.At, 73.21.Ac, 73.22.-f, 73.22.Gk, 81.05.Uw}

\maketitle

%

\section{Introduction}

\label{sec:Introduction}

Graphene, the two dimensional allotrope of carbon, has recently been
attracting a great deal of attention. Since its isolation three years
ago~\citep{NGM+04} a plethora of unusual and interesting properties
has been revealed~\citep{GN07,Kts06rev,NGP+rmp07}. From the point
of view of fundamental physics, low-energy quasi-particles in graphene
behave like massless Dirac fermions propagating at an effective velocity
of light $v\approx10^{6}\,\mbox{ms}^{-2}$. A rather unusual physics
is then observed, where the half-integer quantum Hall effect is a
paradigmatic example~\citep{NGM+05,ZTS+05}. Graphene is also regarded
with great expectations from the point of view of technological applications.
Stability and ballistic transport on the submicrometre scale, even
at room-temperature, make graphene based electronics a promising possibility. 

The possibility of creating stacks of graphene layers with the accuracy
of a single atomic layer, providing an extra dimension to be explored,
is another advantage of graphene for electronic applications. Of particular
interest to us is the double layer of graphene -- the \emph{bilayer}.
Bilayer graphene has shown to have unusual electronic properties,
though unexpectedly dissimilar to those exhibited by its single layer
parent. The new type of integer quantum Hall effect observed in bilayer
graphene \citep{NMcCM+06,MF06}, which is induced by chiral parabolic
bands, is an example of its uniqueness. From the point of view of
applications, bilayer graphene is even more promising for some electronic
devices. It has recently been shown that the band structure of bilayer
graphene can be controlled externally by an applied electric field
so that an electronic gap between the valence and conduction bands
can be tuned in a controllable way \citep{OBS+06,CNM+06,OHL+07}.
This makes the bilayer graphene the only known semiconductor with
a tunable energy gap and may open the door for potential applications
on atomic-scale electronic devices \citep{NNG+06}.

Among the uncommon features of monolayer graphene we find the rather
different behavior of the two possible (perfect) terminations: \emph{zigzag}
and \emph{armchair}. While zigzag edges support localized states,
armchair edges do not \citep{japonese,Dresselhaus,WFA+99}. These
edge states occur at zero energy, the same as the Fermi level of undoped
graphene, meaning that low energy properties may be substantially
altered by their presence. The self-doping phenomenon \citep{PGN06},
the edge magnetization with consequent gap opening in graphene nanoribbons
\citep{SCLprl06}, and half-metallicity \citep{SCLnat06} are examples
of edge states driven effects.

The presence of zero energy edge states at zigzag edges of bilayer
graphene has recently been confirmed assuming a first nearest-neighbor
tight-binding model \citep{CPL+07}. Two families of edge states has
been found to coexist in the bilayer: monolayer edge states, with
finite amplitude on a single plane; and bilayer edge states, with
finite amplitude on both planes, and with an enhanced penetration
into the bulk. As in single layer graphene, bilayer edge states show
up in the electronic spectrum as flat bands at zero energy -- the
Fermi energy of undoped bilayer. These non-dispersive bands gives
rise to a strong peak in the density of states right at the Fermi
energy, which brings about the question of spontaneous magnetic ordering
due to electron-electron interactions.

In the present paper we study the magnetic structure of zigzag bilayer
graphene ribbons induced by electron-electron interactions, which
are included through the Hubbard model. Working within the Hartree
Fock approximation we show that due to the presence of edge states,
which induce a strong peak in the density of states at the Fermi energy,
zigzag bilayer ribbons show edge magnetization even for very small
on-site electronic repulsion. Moreover, it is shown that the spin
configuration is ferromagnetic along the edge, with parallel spins
occurring on both layers, and antiferromagnetic between opposite ribbon
edges. Such a magnetic ordering can be interpreted as being a consequence
of the edge state structure in bilayer graphene.

The paper is organized as follows: in Sec.~\ref{sec:model} we present
the model and the mean field decoupling used here; for a better interpretation
of our results we review briefly in Sec.~\ref{sec:ESNI} the edge
states for non-interacting zigzag bilayer ribbons; in Sec.~\ref{sec:Results}
we present and discuss the results of this work; we close with conclusions
in Sec.~\ref{sec:Conclusions}.

%

\section{Model and mean field treatment}

\label{sec:model}

\begin{figure}
\begin{centering}
\includegraphics[width=1\columnwidth,width=0.9\columnwidth]{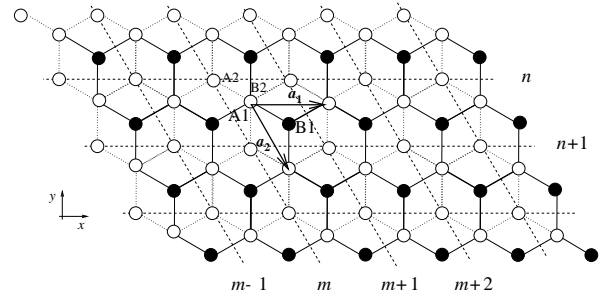}
\par\end{centering}

\caption{\label{cap:ribbon}Ribbon geometry with zigzag edges for bilayer graphene.}

\end{figure}

The study of the magnetic structure in $AB-$stacked bilayer graphene
given here is based on the ribbon geometry with zigzag edges shown
in Fig.~\ref{cap:ribbon}. We use labels~1 and~2 for the top and
the bottom layers, respectively, and labels $Ai$ and $Bi$ for each
of the two sublattices in layer~$i$. Each four-atom unit cell (parallelograms
in Fig.~\ref{cap:ribbon}) has integer indices $m$~(longitudinal)
and $n$~(transverse) such that $m\mathbf{a}_{1}+n\mathbf{a}_{2}$
is its position vector, where $\mathbf{a}_{1}=a(1,0)$ and $\mathbf{a}_{2}=a(1,-\sqrt{3})/2$
are the basis vectors and $a\approx2.46\,\textrm{\AA}$ is the lattice
constant. The simplest model one can write to describe non-interacting
electrons in $AB$-stacked bilayer is the first nearest-neighbor tight-binding
model given by,%
%
\begin{equation}
H_{TB}=\sum_{i=1}^{2}H_{TB,i}+H_{\perp},\label{eq:Htb}\end{equation}
with,%
%
\begin{multline}
H_{TB,i}=-t\sum_{m,n,\sigma}a_{i,\sigma}^{\dagger}(m,n)\big[b_{i,\sigma}(m,n)+b_{i,\sigma}(m-1,n)+\\
b_{i,\sigma}(m,n-1)\big]+\textrm{h.c.},\label{eq:Htbi}\end{multline}
\begin{equation}
H_{\perp}=-t_{\perp}\sum_{m,n,\sigma}a_{1,\sigma}^{\dagger}(m,n)b_{2,\sigma}(m,n)+\textrm{h.c.},\label{eq:Hperp}\end{equation}
where $a_{i,\sigma}(m,n)$ {[}$b_{i,\sigma}(m,n)$] is the annihilation
operator for the state in sublattice $Ai$ ($Bi$), $i=1,2$, at position
($m,n$), and spin $\sigma=\uparrow,\downarrow$. The first term on
the right hand side of Eq.~(\ref{eq:Htb}) describes in-plane hopping,
$t\approx2.7\,\textrm{eV}$, while the second term parametrizes the
inter-layer coupling, $t_{\perp}/t\ll1$. In order to examine the
magnetic polarization due to electron-electron interactions we add
the Hubbard term to Eq.~\ref{eq:Htb}. The total Hamiltonian describing
the bilayer system reads,%
%
\begin{equation}
H=H_{TB}+H_{U},\label{eq:H}\end{equation}
where $H_{U}$ represents the on-site Coulomb interaction,%
%
\begin{align}
H_{U}=U\sum_{i=1}^{2} & \sum_{m,n}\big[a_{i,\uparrow}^{\dagger}(m,n)a_{i,\uparrow}(m,n)a_{i,\downarrow}^{\dagger}(m,n)a_{i,\downarrow}(m,n)\nonumber \\
 & +b_{i,\uparrow}^{\dagger}(m,n)b_{i,\uparrow}(m,n)b_{i,\downarrow}^{\dagger}(m,n)b_{i,\downarrow}(m,n)\big],\label{eq:HU}\end{align}
The Hubbard model is a good starting point to study magnetism whenever
the density of states at the Fermi energy is large enough to produce
effective screening of the Coulomb interaction. This is true for the
clean bilayer, where a finite density of states at the neutrality
point produces some amount of screening in the system~\citep{WC06}.
It is certainly the case in the presence of zigzag edges, where the
density of states peak at the Fermi energy implies very effective
screening.

The system Hamiltonian in Eq.~(\ref{eq:H}) is treated here within
mean-field theory. In the Hartree Fock approximation the mean-field
version of Eq.~(\ref{eq:H}) reads,%
%
\begin{equation}
H_{MF}=H_{TB}+H_{U}^{MF},\label{eq:Hhf}\end{equation}
with%
%
\begin{align}
H_{U}^{MF}=U\sum_{i=1}^{2} & \sum_{m,n,\sigma}\big[\tilde{n}_{Ai,-\sigma}(m,n)a_{i,\sigma}^{\dagger}(m,n)a_{i,\sigma}(m,n)\nonumber \\
 & +\tilde{n}_{Bi,-\sigma}(m,n)b_{i,\sigma}^{\dagger}(m,n)b_{i,\sigma}(m,n)\big],\label{eq:HUhf}\end{align}
where $\tilde{n}_{\Gamma i,\sigma}(m,n)$ is the electronic density
for spin $\sigma=\uparrow,\downarrow$ at the site of sublattice $\Gamma=A,B$
and layer $i=1,2$ of the cell $(m,n)$. The electronic spin densities
$\tilde{n}_{\Gamma i,\sigma}(m,n)$ have to be determined self-consistently
through,%
%
\begin{align}
\tilde{n}_{Ai,\sigma}(m,n)= & \left\langle a_{i,\sigma}^{\dagger}(m,n)a_{i,\sigma}(m,n)\right\rangle _{MF},\label{eq:spdA}\\
\tilde{n}_{Bi,\sigma}(m,n)= & \left\langle b_{i,\sigma}^{\dagger}(m,n)b_{i,\sigma}(m,n)\right\rangle _{MF},\label{eq:spdB}\end{align}
where the average $\left\langle \cdots\right\rangle _{MF}$ is done
with the mean-field Hamiltonian in Eq.~(\ref{eq:Hhf}). Quantum fluctuations,
which are ignored within mean-field theory, are expected to reduce
the magnetic moments but not to change significantly the overall magnetic
structure. As a further approximation we assume that the self-consistent
solution of Eqs.~(\ref{eq:spdA}) and~(\ref{eq:spdB}) is $m$ independent,
i.e.,\begin{align}
\tilde{n}_{Ai,\sigma}(m,n)\equiv & \tilde{n}_{Ai,\sigma}(n)=\frac{1}{L}\sum_{m}\tilde{n}_{Ai,\sigma}(m,n),\label{eq:spdAap}\\
\tilde{n}_{Bi,\sigma}(m,n)\equiv & \tilde{n}_{Bi,\sigma}(n)=\frac{1}{L}\sum_{m}\tilde{n}_{Bi,\sigma}(m,n),\label{eq:spdBap}\end{align}
where $L$ is the longitudinal ribbon length. We can justify this
approximation here because we are mainly interested on the study of
edge magnetization when edge states are present, and, as we will see
in Sec.~\ref{sec:ESNI}, edge states are homogeneous along the edge.
Note, however, that we keep the sublattice index in Eqs.~(\ref{eq:spdAap})
and~(\ref{eq:spdBap}), meaning that we can still have in-cell inhomogeneity.

Without loss of generality we assume that the ribbon in Fig.~\ref{cap:ribbon}
has $N$ unit cells in the transverse cross section ($y$ direction)
with $n\in\{0,\dots,N-1\}$, and we use periodic boundary conditions
along the longitudinal direction ($x$ direction). Noting the translational
invariance of the ribbon along the $x$ direction, and having Eqs.~(\ref{eq:spdAap})
and~(\ref{eq:spdBap}) in mind, it is easy to diagonalize Hamiltonian~(\ref{eq:Hhf})
with respect to the $m$~index just by Fourier transform along the
longitudinal direction, $H=\sum_{k}\, H_{k}$, with $H_{k}$ given
by,%
%
\begin{equation}
H_{k}=H_{TB,k}+H_{U,k}^{MF},\label{eq:Hk}\end{equation}
where,%
%
\begin{multline}
H_{TB,k}=\\
-t\sum_{i=1}^{2}\sum_{n,\sigma}a_{i,\sigma}^{\dagger}(k,n)[(1+e^{ik})b_{i,\sigma}(k,n)+b_{i,\sigma}(k,n-1)]\\
-t_{\perp}\sum_{n,\sigma}a_{1,\sigma}^{\dagger}(k,n)b_{2,\sigma}(k,n)+\textrm{h.c.}\,,\label{eq:Htbk}\end{multline}
and,%
%
\begin{align}
H_{U,k}^{MF}=U\sum_{i=1}^{2} & \sum_{n,\sigma}\big[\tilde{n}_{Ai,-\sigma}(n)a_{i,\sigma}^{\dagger}(k,n)a_{i,\sigma}(k,n)\nonumber \\
 & +\tilde{n}_{Bi,-\sigma}(n)b_{i,\sigma}^{\dagger}(k,n)b_{i,\sigma}(k,n)\big],\label{eq:HUhfk}\end{align}
with self-consistent spin densities given by Eqs.~(\ref{eq:spdAap})
and~(\ref{eq:spdBap}), which can be rewritten as,%
%
\begin{align}
\tilde{n}_{Ai,\sigma}(n)= & \frac{1}{L}\sum_{k}\left\langle a_{i,\sigma}^{\dagger}(k,n)a_{i,\sigma}(k,n)\right\rangle _{MF},\label{eq:spdAk}\\
\tilde{n}_{Bi,\sigma}(n)= & \frac{1}{L}\sum_{k}\left\langle b_{i,\sigma}^{\dagger}(k,n)b_{i,\sigma}(k,n)\right\rangle _{MF}.\label{eq:spdBk}\end{align}
All conclusions presented in Sec.~\ref{sec:Results} regarding the
magnetic structure of zigzag bilayer ribbons are drawn by solving
Eqs.~(\ref{eq:Hk}-\ref{eq:spdBk}).

%

\section{Edge states in the non-interacting limit}

\label{sec:ESNI}

It is shown in Sec.~\ref{sec:Results} that the results for the edge
magnetization of zigzag bilayer ribbons are a consequence of the edge
state structure found in this system~\citep{CPL+07}. In this section
we briefly review the main features of bilayer edge states for $U=0$
in Eq.~(\ref{eq:H}), i.e., in the absence of interactions.

\begin{figure}
\begin{centering}
\includegraphics[width=0.9\columnwidth]{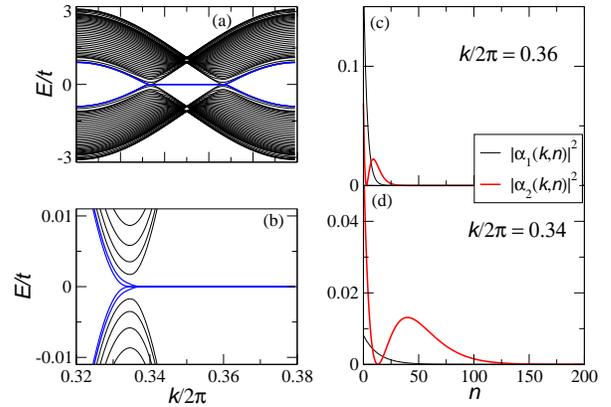}
\par\end{centering}

\caption{\label{fig:ESnoU}(Color online) (a) - Energy spectrum for a graphene
bilayer ribbon with zigzag edges for $N=400$. (b) - Zoom in of panel
(a). (c) - Charge density of the edge states at $k/2\pi=0.36$. (d)
- The same as in (c) at $k/2\pi=0.364$. The interlayer coupling was
set to $t_{\perp}/t=0.2$ in all panels.}

\end{figure}

The band structure of a bilayer ribbon with zigzag edges is shown
in Fig.~\ref{fig:ESnoU}~(a) for $N=400$, obtained by numerically
solving Eq.~(\ref{eq:Htbk}). We can see the partly flat bands at
$E=0$ for $k$ in the range $2\pi/3\leq ka\leq4\pi/3$, corresponding
to four edge states, two per edge. The zoom shown in Fig.~\ref{fig:ESnoU}~(b)
for $ka\approx2\pi/3$ clearly shows that there are four flat bands. 

In order to understand the spatial structure of edge states in bilayer
graphene we solve the Schrödinger equation, $H_{TB,k}\left|\mu,k\right\rangle =E_{\mu,k}\left|\mu,k\right\rangle $,
for $E_{\mu,k}=0$, where $\mu$ labels the eigenstate index including
spin. First we note that Hamiltonian $H_{TB,k}$ in Eq.~(\ref{eq:Htbk})
effectively defines a 1D problem in the transverse direction of the
ribbon. It is then possible to write any eigenstate $\left|\mu,k\right\rangle $
as a linear combination of the site amplitudes along the cross section,%
%
\begin{multline}
\left|\mu,k\right\rangle =\\
\sum_{n}\sum_{i=1}^{2}\big[\alpha_{i}(k,n)\left|a_{i},k,n,\sigma\right\rangle +\beta_{i}(k,n)\left|b_{i},k,n,\sigma\right\rangle \big],\label{eq:eigstate}\end{multline}
where the four terms per~$n$ refer to the four atoms per unit cell,
to which we associate the one-particle states $\left|c_{i},k,n,\sigma\right\rangle =c_{i,\sigma}^{\dagger}(k,n)\left|0\right\rangle $,
with $c_{i,\sigma}=a_{i,\sigma},b_{i,\sigma}$, spin $\sigma=\uparrow,\downarrow$,
and $i=1,2$. To account for the finite width of the ribbon we require
the following boundary conditions,%
%
\begin{equation}
\alpha_{1}(k,N)=\alpha_{2}(k,N)=\beta_{1}(k,-1)=\beta_{2}(k,-1)=0.\label{eq:BC}\end{equation}
After solving the Shrödinger equation for zero energy and the boundary
conditions in Eq.~(\ref{eq:BC}) we find four possible eigenstates
per $k$, where the only nonzero coefficients for each of them are
given by~\citep{CPL+07}:%
%
\begin{align}
\alpha_{1}(k,n)=0\,,\hspace{0.5cm} & \alpha_{2}(k,n)=\alpha_{2}(k,0)D_{k}^{n}e^{-i\frac{ka}{2}n}\,;\label{eq:sol1A}\end{align}
\begin{equation}
\begin{array}{l}
\alpha_{1}(k,n)=\alpha_{1}(k,0)D_{k}^{n}e^{-i\frac{ka}{2}n},\\
\alpha_{2}(k,n)=-\alpha_{1}(k,0)D_{k}^{n-1}\frac{t_{\perp}}{t}e^{-i\frac{ka}{2}(n-1)}\Big(n-\frac{D_{k}^{2}}{1-D_{k}^{2}}\Big);\end{array}\label{eq:sol2A}\end{equation}
\begin{align}
\beta_{1}(k,n)=\beta_{1}(k,N-1)D_{k}^{n'}e^{i\frac{ka}{2}n'},\hspace{0.5cm} & \beta_{2}(k,n)=0\,;\label{eq:sol1B}\end{align}
and\begin{equation}
\begin{array}{l}
\beta_{1}(k,n)=-\beta_{2}(k,N-1)D_{k}^{n'-1}\frac{t_{\perp}}{t}e^{i\frac{ka}{2}(n'-1)}\Big(n'-\frac{D_{k}^{2}}{1-D_{k}^{2}}\Big),\\
\beta_{2}(k,n)=\beta_{2}(k,N-1)D_{k}^{n'}e^{i\frac{ka}{2}n'}\,;\end{array}\label{eq:sol2B}\end{equation}
where $D_{k}=-2\cos(ka/2)$ and $n=N-n'-1$, with $n'\in\{0,\dots,N-1\}$.
As is easily seen, the coefficients in Eqs.~(\ref{eq:sol1A}-\ref{eq:sol2B})
give convergent wave functions only if $2\pi/3<ka<4\pi/3$, in which
case they represent zero energy states localized at the surface~--~edge
states~--~and provide an explanation for the four flat zero energy
bands in Fig.~\ref{fig:ESnoU}~(a) and~(b). Equations~(\ref{eq:sol1A})
and~(\ref{eq:sol2A}) correspond to edge states localized at the
top zigzag edge in Fig.~\ref{cap:ribbon}, and Eqs.~(\ref{eq:sol1B})
and~(\ref{eq:sol2B}) are edge states localized at the bottom zigzag
edge in the same figure. Note, however, that the solutions given by
Eqs.~(\ref{eq:sol1A}-\ref{eq:sol2B}) are exact eigenstates only
for semi infinite systems, where the boundary conditions given in
Eq.~(\ref{eq:BC}) are fully satisfied. In a finite ribbon overlapping
of the four edge states leads to a slight dispersion and non-degeneracy.
Nevertheless, as long as the ribbon width is sufficiently large, this
effect is only important at $ka\simeq2\pi/3$ and $ka\simeq4\pi/3$
where the localization length is large enough for the overlapping
to be appreciable~\citep{WFA+99}. For completeness we give the normalization
constants appearing in Eqs.~(\ref{eq:sol1A}-\ref{eq:sol2B}),%
%
 \begin{align}
|\alpha_{2}(k,0)|^{2}=|\beta_{1}(k,N-1)|^{2} & =1-D_{k}^{2},\label{eq:normconst1}\\
|\alpha_{1}(k,0)|^{2}=|\beta_{2}(k,N-1)|^{2} & =\frac{(1-D_{k}^{2})^{3}}{(1-D_{k}^{2})^{2}+t_{\perp}^{2}/t^{2}}.\label{eq:normconst2}\end{align}

An example of the charge density associated with Eq.~(\ref{eq:sol2A})
is shown in panels~(c) and~(d) of Fig.~\ref{fig:ESnoU} for $t_{\perp}/t=0.2$,
where the $|\alpha_{1}(k,n)|^{2}$ dependence can also be seen as
the solution given by Eq.~(\ref{eq:sol1A}) for $|\alpha_{2}(k,n)|^{2}$,
apart from a normalization factor. Of particular interest to understand
the magnetic structure due to interaction effects is the fact that
edge states in zigzag bilayer graphene are such that at one edge they
live only on sublattice~$A$ whereas at the opposite edge they live
on sublattice~$B$.

%

\section{Results and discussion}

\label{sec:Results}

\begin{figure}
\begin{centering}
\includegraphics[width=0.9\columnwidth]{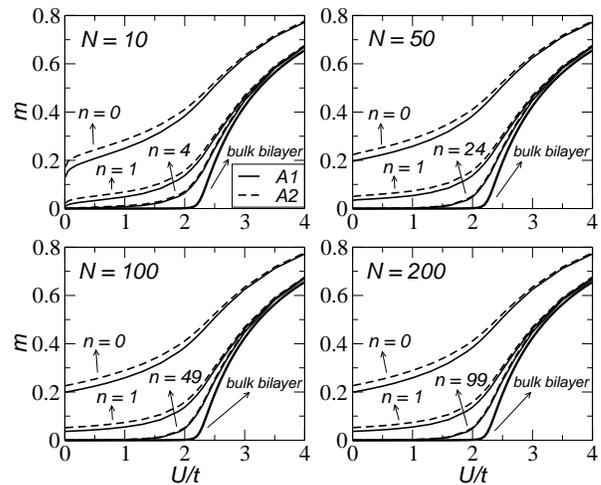}
\par\end{centering}

\caption{\label{fig:magU} Dependence of the magnetization $m=\tilde{n}_{Ai,\uparrow}-\tilde{n}_{Ai,\downarrow}$,
with $i=1,2$, on the interaction parameter $U$ for different ribbon
widths $N$. The magnetization was computed at sites $n=0$, $n=1$,
and at the middle of the ribbon. Solid lines are for the upper layer
($i=1$) and dashed lines for the bottom layer ($i=2$). The result
for graphite double sheet (bulk bilayer) is also shown.}

\end{figure}

In Fig.~\ref{fig:magU} the results for the local magnetization $m=\tilde{n}_{Ai,\uparrow}-\tilde{n}_{Ai,\downarrow}$,
for $i=1,2$, are shown as a function of the Hubbard parameter $U$
for different ribbon widths~$N$. For each ribbon width we have computed
the local magnetization at sites of the $A$ sublattice belonging
to cells $n=0$, $n=1$, and right at the middle of the ribbon (see
Fig.~\ref{cap:ribbon}). The first conclusion we can draw is that
sites near the edge get polarized even for very small $U$, while
sites in the middle of the ribbon behave like bulk bilayer~\citep{foot1}.
Another interesting feature shown in Fig.~\ref{fig:magU} is that
at the considered edge the magnetization of $A2$ sites is larger
than that of $A1$ sites, an asymmetry that vanishes away from the
edge. We will come back to this below. As regards the $B$ sublattice
its magnetization (not shown in Fig.~\ref{fig:magU}) is always similar
to the bulk result even right at the edge ($n=0$). However, when
we move to the opposite edge, the $A$ and $B$ sublattices change
roles: $B$ sites at the opposite edge get polarized for very small
$U$ while $A$ sites show the bulk result. The conclusion then is
that edge magnetization involving different sublattices at opposite
edges is showing up in zigzag bilayer ribbons, even for very small
$U$. In particular we get $m\approx0.2\mu_{B}$ right at the edge
for $U=0.1t\approx0.3\,\mbox{eV}$, similar to what is found in graphene~\citep{japonese}.

\begin{figure}
\begin{centering}
\includegraphics[width=0.9\columnwidth]{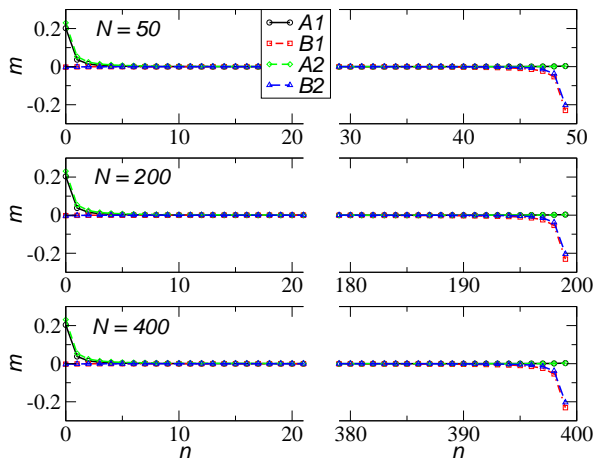}
\par\end{centering}

\caption{\label{fig:magn}(Color online) Magnetization $m=\tilde{n}_{\Gamma i,\uparrow}-\tilde{n}_{\Gamma i,\downarrow}$
along the ribbon cross section for the $\Gamma i=A1,A2,B1,B2$. Three
different ribbon widths were considered: $N=50,200,400$ from top
to bottom. The interaction parameter was set to $U=0.1t$.}

\end{figure}

A better understanding of the edge magnetization is achieved by fixing
$U$ and plotting the local magnetization $m=\tilde{n}_{\Gamma i,\uparrow}-\tilde{n}_{\Gamma i,\downarrow}$
across the ribbon section for $\Gamma i=A1,A2,B1,B2$. This is done
in Fig.~\ref{fig:magn} for different ribbon widths and for a fixed
interaction parameter $U=0.1t$. As is clearly seen, for such a small
interaction only the edges are polarized. Moreover, the edge magnetization
is opposite on opposite edges~--~\emph{antiferromagnetic arrangement
across the ribbon}. Also, we can see that at the edge starting with
cell $n=0$ only sublattice~$A$ has a finite magnetization, whereas
at the opposite edge only sublattice~$B$ has non-vanishing magnetization.
Finally, it is also apparent that at each edge the non-zero sublattice
magnetization has same sign in both layers~--~\emph{ferromagnetic
arrangement along the edge}. These observations are consistent with
first-principles density-functional calculations of the magnetic structure
of graphitic fragments (infinite number of layers) \citep{LSP+05}
and bilayer graphene nanoribbons \citep{SMD+08}.

We have seen in Sec.~\ref{sec:ESNI} that bilayer edge states have
the following property: at the edge starting with $n=0$ they live
only on sublattice~$A$, while at the opposite edge they live only
on sublattice~$B$, as given by Eqs.~(\ref{eq:sol1A}-\ref{eq:sol2B}).
The above results for the edge magnetization may therefore be attributed
to the polarization of edge states in order to reduce on-site Coulomb
energy. This interpretation also provides an explanation for the layer
difference in local magnetization. As mentioned before, it can be
seen in Fig.~\ref{fig:magU} that the magnetization at $A2$ sites
is higher than at $A1$ sites for the edge starting with $n=0$. If
we recall Eqs.~(\ref{eq:sol1A}) and~(\ref{eq:sol2A}) for the wave
function amplitudes at the considered edge we immediately see that
while the two edge state families contribute to $A2$ only one has
finite amplitude at $A1$ sites. The same is true for $B1$ and $B2$
sites, in agreement with Eqs.~(\ref{eq:sol1B}) and~(\ref{eq:sol2B}).
As regards the antiferromagnetic polarization between edge states
living in opposite edges, it guarantees a ground state with zero total
magnetization, as it is known to be the case for the half-filled Hubbard
model~\citep{Lieb89}.

Finally we note that edge magnetization gives rise to a finite gap
at the Fermi level, in complete analogy to monolayer graphene~\citep{SCLprl06}.
Half-metallicity has been predicted for zigzag single layer ribbons
due to the edge magnetization and the presence of a finite gap~\citep{SCLnat06}.
We expect that bilayer ribbons also become half-metallic, with an
extra switching capability owing to the effect of a perpendicular
electric field~\citep{CNM+06,OHL+07}.

%

\section{Conclusions}

\label{sec:Conclusions} 

We have studied the edge magnetization in bilayer graphene ribbons
with zigzag edges. The presence of flat edge-state bands at the Fermi
energy of undoped bilayer, which gives rise to a strong peak in the
density of states, makes bilayer ribbons magnetic at the edges even
for very small on-site electronic repulsion. Using the Hubbard model
in the Hartree Fock approximation we have shown that the magnetic
structure in bilayer ribbons with zigzag edges is ferromagnetic along
the edge, involving sites of the two layers but belonging to the same
sublattice, and antiferromagnetic between opposite edges and involving
sites of different sublattices. This magnetic structure is a consequence
of the nature of the edge states present in bilayer ribbons with zigzag
edges.

The experimental observation of edge magnetism in bilayer graphene
nanoribbons, and possible application as graphene-based magnetic nanostructures,
has the 1D nature of the spin polarized state as a major drawback.
In single layer graphene a crossover temperature $T_{x}\approx10\,\mbox{K}$
has recently been estimated for the magnetic correlations at zigzag
edges \citep{YK07}. Below $T_{x}$ the spin correlation length grows
exponentially with decreasing temperature, while above $T_{x}$ it
is inversely proportional to the temperature. This behavior limits
the long-range magnetic order to $\sim1\,\mbox{nm}$ at $300\,\mbox{K}$.
We have shown, within Hartree Fock, that zigzag bilayer nanoribbons
have a broken-symmetry ground state with a finite spin polarization
along the edges even at small on-site Hubbard repulsion. The presence
of an extra layer with respect to the monolayer case, and the fact
that bilayer edge states have an enhanced penetration into the bulk,
should affect the crossover temperature $T_{x}$ and the room temperature
correlation length. Further work is needed to understand to what extent.


\subsection*{Acknowledgments}

E.V.C. acknowledges the financial support of Fundação para a Ciência
e a Tecnologia through Grant No.~SFRH/BD/13182/2003. E.V.C., N.M.R.P.,
and J.M.B.L.S. acknowledge financial support from POCI~2010 via project
PTDC/FIS/64404/2006.



\bibliographystyle{apsrev}

\bibliographystyle{apsrev}
\bibliography{Bibliography/graphenetheo,Bibliography/grapheneexp,Bibliography/recursionMethod,Bibliography/condmattbooks,Bibliography/myrefs,Bibliography/footnotes,Bibliography/graphenerev}

\end{document}